# A Quantum Phase Neural Network with Multi-Inputs/Single-Output

Shuang Cong, Jinmin Yang, Sajede Harraz
*Dept, of Automation, University of Science and Technology of China*
Hefei China
scong@ustc.edu.cn

***Abstract*—A 35 input/single output quantum phase neural network (QPNN) is proposed to recognize the English letters' A 'and' B 'composed $7 \times 5 = 35$ of pixels. Each node of the proposed QPNN has the input with 0 or 1. Through normalization and quantum phase encoding, each set of 35 with 0 or 1 digital inputs is encoded into a quantum superposition state described by 35 quantum phases. Then, a quantum rotation gate is used to introduce adjustable phase, and a controlled NOT gate is used to entangle the relative phase between the two inputs. The relationship between the input and output of the network is derived. This paper also derived the phase weight learning training algorithm with adaptive learning rate. The solutions with the recognition probability of 1 is designed, and the analytical expressions for all adjustable phase solutions with zero errors are not unique. The experimental performance is verified on the Qiskit platform. This paper uses phase drive to convert the probability calculation into an analytical polynomial function with adjustable rotation angles in the network, avoiding the bottleneck of generating all $2^{35}$ complex exponential amplitudes and providing an effective way to solve the "exponential wall" problem, which provides a new implementation solution for the practical application of multi-input recognition problems.**

***Keywords—quantum phase neurons, quantum phase neural network, quantum neural networks, multi-inputs/single-output***

## I. INTRODUCTION

The concept of quantum neural networks was first proposed by Kak in 1995 [1], which opened the way for combining neural networks with quantum computing to form a new computing paradigm. In 1998, Menneer conducted a comprehensive comparison of the performance of classical neural networks and quantum neural networks [2], proving that the performance of quantum neural networks is superior to that of classical neural networks. In 2000, Behrman et al. proposed a quantum dot neural network model [3], which proved that a single real-time quantum dot molecule can be used as a recursive time QNN. In 2005, Noriaki Kouda et al. proposed a quantum neuron based on the quantum phase [4] and used it as a basis to construct a QNN, achieving a working mode similar to that of classical neural networks. In 2014, Schuld et al. systematically analyzed QNN models [5] and summarized three essential conditions for building a quantum neural network: the input and output of the network must be binary strings encoded as quantum states, the network must reflect the basic mechanisms of classical neural network operations, and the training and iteration of the network must conform to quantum properties such as maintaining unitary properties.

Although the QNN has shown initial application potential in several practical problems, designing and implementing an n-bit system remains a challenge for people, despite its ability to store $2^n$-dimensional parameters. The amplitude encoding and output projection measurements are currently used to encode all $2^n$ parameters in an $n$-qubit system and perform quantum operator operations. When projection measurements are used to determine the quantum state, the output of the $n$-qubit system needs to undergo $2^n$ exponential measurements to obtain the output value of the model. The computational complexity in its model increases exponentially with increasing number of qubits n, indicating the existence of an "exponential wall" problem, which leads to an exponential increase in the computation time of the model in high-qubit systems.

In view of the above challenges, this paper proposes a novel quantum neural network architecture based on the quantum phase. All operations in the network are implemented on the basis of basic quantum logic gates, ensuring that the overall process strictly conforms to the characteristics of quantum mechanics. The calculation of the output quantum state in the $2^n$ dimensional Hilbert space composed of $n$ qubits is transformed into the calculation of the polynomial of the output quantum state phase through phase encoding and operation, thereby avoiding the exponential growth of the output quantum state calculation with increasing number of quantum bits n and truly demonstrating the advantages of quantum systems. Moreover, this article also proposes a multi-input/single-output quantum phase neural network (QPNN), providing the entire process of designing the QPNN structure, as well as the derivation process of the adaptive correction formula for weighted phase training.

The structure of this paper is as follows: Section II presents the design of a multi-input/odd-output all phase quantum model, Section III describes the construction of a quantum circuit structure diagram based on the multi-input/single-output QPNN structure diagram, Section IV presents the derivation of the network input/output relationship, Section V presents the derivation of the network weight phase training formula, Section VI presents the simulation experiment of letter recognition and the experimental verification based on the Qiskit platform, and Section VII presents a summary of this paper.

The work is supported by National Natural Science Foundation of China under Grant 62473354.

## II. Design of the multi-imput/single-output QPNN model

The multi-input/single-output quantum phase neural network (QPNN) created is composed of multiple quantum phase neurons [6] connected in parallel. The most basic one is the QPNN for image recognition, which is composed of one QPP. The structure of the QPNN model is shown in Figure 1: 1) Each image is digitized into a Boolean matrix of $r \times l$ and then converted into a column vector of $r \times l = m$ as a set of input variables; when the output of the model is a single node, its output is one of two integers 0 or 1, representing different input patterns. At this time, the single-output model can recognize two different images or patterns; 2) because the initial image may be binary {0,1}, 0-1 quantum state, or any real number, any QPNN must include normalization of the input data and encoding of the quantum phase; 3) afterwards, according to the required structure, the same $n$ quantum phase neurons are used, which are a network of quantum gate operations consisting of m input nodes and a single output node. 4) The output of the entire QPNN is the expected output [0,1].

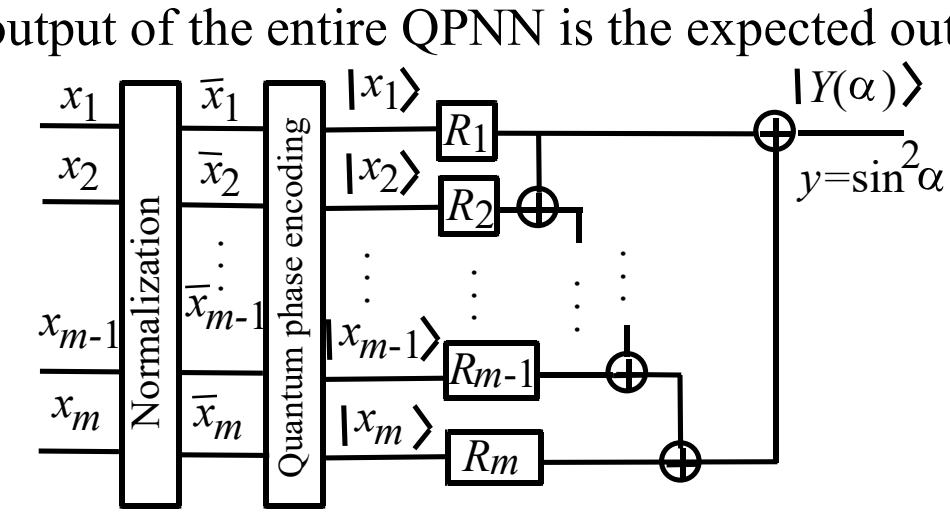

Fig. 1 Network structure diagram of m-input/single-output QPNN

In this paper, the recognition of English letters A and B is taken as an example to describe the design process of m input/single-output QPNN network structures. The recognition of the English letters A and B for $7\times 5=35$ pixels is realized. Each letter is digitized into a Boolean matrix of, for example, 7 × 5, and then it is converted into a column vector of $m = 7 \times 5 = 35$ as a set of input variables (nodes). In this way, two different types of input data for letters A and B are obtained, and the corresponding numbers of labels for the expected node outputs are 0 and 1, respectively. On this basis, the QPNN designed for recognizing 2 letters consists of $m$=35 with real-valued $\{0,1\}$ inputs $X=(x_1,x_2,\ldots,x_{35})^{\mathrm{T}}$ and 1 real-valued output.

The normalization formula for input variables $\bar{X}=(\bar{x}_1,\bar{x}_2,\ldots,\bar{x}_{35})^T$ with real values $\{0,1\}$ is as follows:

$$\bar{x}_j=\frac{x_j}{\sqrt{\sum_{j=1}^{m}x_j^2}} \tag{1}$$

where $\bar{x}_j$ is the normalized input variable, $j$ =1, 2,…,$m$ ($m$=35).

The normalized input variables $\bar{X}=(\bar{x}_1,\bar{x}_2,\ldots,\bar{x}_{35})^T$ need to be transformed into quantum state forms described by phase through quantum phase encoding. We use a phase encoding method to convert the normalized input variables $\bar{X}=(\bar{x}_1,\bar{x}_2,\ldots,\bar{x}_{35})^T$ with $\{0,1\}$ real values into quantum states and define the following conversion formula to encode real input into a quantum state described by phase $|x_j\rangle$:

$$\begin{aligned}|x_j\rangle &= \cos\left[\frac{\pi}{2}\left(\frac{\bar{x}_j-\min(\bar{X})}{\max(\bar{X})-\min(\bar{X})}\right)\right]|0\rangle+\sin\left[\frac{\pi}{2}\left(\frac{\bar{x}_j-\min(\bar{X})}{\max(\bar{X})-\min(\bar{X})}\right)\right]|1\rangle \\ &= \begin{pmatrix}\cos\left[\frac{\pi}{2}\left(\frac{\bar{x}_j-\min(\bar{X})}{\max(\bar{X})-\min(\bar{X})}\right)\right] \\ \sin\left[\frac{\pi}{2}\left(\frac{\bar{x}_j-\min(\bar{X})}{\max(\bar{X})-\min(\bar{X})}\right)\right]\end{pmatrix}\end{aligned} \tag{2}$$

where the phase of the quantum state $|x_j\rangle$ is $\frac{\pi}{2}\left(\frac{\bar{x}_j-\min(\bar{X})}{\max(\bar{X})-\min(\bar{X})}\right)$.

Next, one needs to take the design process of a multi-input/single-output QPNN to train and recognize the letters A and B as an example to describe in detail. After normalization and quantum encoding, the obtained quantum states are input into a quantum manipulator for recognition. This quantum manipulator has only one real output y and uses expected outputs of 0 and 1 to recognize the input letters A and B. In this paper, the expected output is set as follows: the expected output for the letter A is 0, and the expected output for the letter B is 1.

## III Construction of the Quantum Citcuit Diagram

The quantum manipulator is implemented using parameterized quantum circuits, and the quantum circuit diagram of the $i$-th quantum manipulator is shown in Figure 2. The first output of the $i$-th quantum manipulator is the only external output of the quantum manipulator.

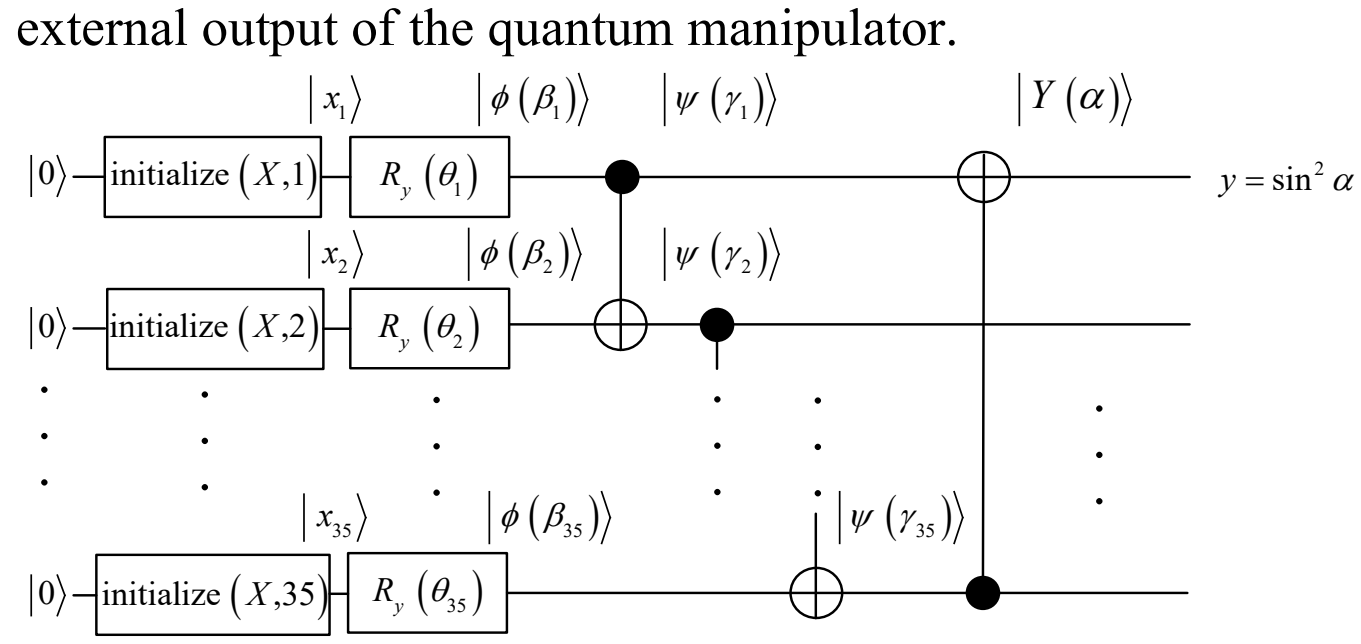

Fig. 2 Quantum circuit diagram of the quantum manipulator

According to Figure 2, the working process of the designed quantum manipulator can be described as follows: 1) Quantum state initialization, including normalization processing and quantum phase encoding, implemented by the designed **initialize** function; 2) weighted phase deviation, achieved by a $R_y$ rotation gate; and 3) realization of phase summation through controlled NOT gates. Quantum state initialization is the process of converting the default state $|0\rangle$ when a quantum circuit is created to the target quantum state through specific operations to participate in subsequent operations. In the Qiskit simulation experiment, this paper uses the **initialize** function to implement this function. The real number input and index *j* are input into the initialize function. The function is normalized, and the quantum state is encoded according to formulas (1) and (2) to calculate the target quantum state $\left|x_j\right\rangle$ described by the phase. Afterward, the created *j*-th state quantum state can be converted into the target state through index *j*. Therefore, in Figure 2, initialize (*X, j*) is used to represent the quantum state initialization operation. The function "initialize" can also be described using the following formula:

$$\left|0\right\rangle \xrightarrow{\text{initialize}(X,\ j)} \left|x_j\right\rangle \tag{3}$$

where the quantum state $\left|x_j\right\rangle$ is obtained by Qiskit code by operating on the input variables *X* according to (1) and (2).

Specifically, by inputting the real number and index *j* into the initialize function, the initialize function first normalizes and encodes the quantum state according to equations (1) and (2), calculates the target quantum state described by the phase, and then inputs the target quantum state $\left|x_j\right\rangle$ and index *j* into the Qiskit platform's built-in quantum state initialization function to convert the *j*-th quantum state $\left|0\right\rangle$ into the target state $\left|x_j\right\rangle$. Therefore, initialize (*X, j*) is used in Figure 3 to represent the quantum state initialization operation.

The rotation operation of the weighted phase is achieved through a phase rotation gate $R_y(\theta_j)$ to achieve phase deviation, with a deviation angle of $\theta_j/2$. The input quantum state $\left|x_j\right\rangle$ is deflected by the phase shift gate to obtain a new quantum state $\left|\phi\left(\beta_j\right)\right\rangle$ with a phase of $\beta_j$.

After a new quantum state $\left|\phi\left(\beta_j\right)\right\rangle$ is obtained, a controlled NOT gate is used for phase summation. First, a quantum state $\left|\phi\left(\beta_1\right)\right\rangle$ is used as a control bit to control the controlled NOT gate to act on the quantum state $\left|\phi\left(\beta_2\right)\right\rangle$ for phase summation. After the action, the quantum state $\left|\psi\left(\gamma_1\right)\right\rangle$ is represented by a quantum state ( $\left|\phi\left(\beta_1\right)\right\rangle = \left|\psi\left(\gamma_1\right)\right\rangle$ , and $\beta_1 = \gamma_1$ ), and the quantum state $\left|\phi\left(\beta_2\right)\right\rangle$ is summed in phase to obtain a new quantum state $\left|\psi\left(\gamma_2\right)\right\rangle$ with a phase of $\gamma_2$. Then, the quantum state $\left|\psi\left(\gamma_2\right)\right\rangle$ is used as a control bit to control the controlled NOT gate to act on the quantum state $\left|\phi\left(\beta_3\right)\right\rangle$ to obtain a new quantum state $\left|\psi\left(\gamma_3\right)\right\rangle$ . This process continues until the quantum state $\left|\psi\left(\gamma_{35}\right)\right\rangle$ is obtained.

To obtain network output $\left|Y(\alpha)\right\rangle$ , the quantum state $\left|\psi\left(\gamma_{35}\right)\right\rangle$ needs to be used as a control bit to control the controlled NOT gate to act on the quantum state $\left|\psi\left(\gamma_1\right)\right\rangle$ for phase summation, which can obtain the expression of the output quantum state $\left|Y(\alpha)\right\rangle$. By measuring the quantum state, the phase $\alpha$ can be obtained, and the calculated value $\sin^2\alpha$ can be used to obtain the probability $y = \sin^2\alpha$ of the quantum state $\left|Y(\alpha)\right\rangle$ in the quantum basis state $\left|1\right\rangle$ as the real output of the quantum manipulator.

## IV. Drivation of the QPNN Input/Output Relationships

### *A. Deduction of the Weight Phase Deviation Formula*

The effect of an adjustable weight phase on quantum states is achieved through phase rotation gates. First, choose the phase shift gate. The commonly used phase rotation operators in quantum computing are $R_x$, $R_y$, and $R_z$. In this paper, the gate $R_y\left(\theta_j\right)$ ( $j = 1, 2, ..., 35$ ) is chosen as the weight phase function.

The new quantum state $\left|\phi\left(\beta_j\right)\right\rangle$ obtained by the deflection angle $\theta_j/2$ of the input quantum state $\left|x_j\right\rangle$ under the action of a phase rotation gate $R_y\left(\theta_j\right)$ is as follows:

$$\begin{aligned}
\left|\phi(\beta_j)\right\rangle &= R_y(\theta_j)\left|x_j\right\rangle \\
&= \begin{pmatrix} \cos\frac{\theta_j}{2} & -\sin\frac{\theta_j}{2} \\ \sin\frac{\theta_j}{2} & \cos\frac{\theta_j}{2} \end{pmatrix} \begin{pmatrix} \cos\left[\frac{\pi}{2}\left(\frac{\bar{x}_j - \min(\bar{X})}{\max(\bar{X}) - \min(\bar{X})}\right)\right] \\ \sin\left[\frac{\pi}{2}\left(\frac{\bar{x}_j - \min(\bar{X})}{\max(\bar{X}) - \min(\bar{X})}\right)\right] \end{pmatrix} \\
&= \begin{pmatrix} \cos\frac{\theta_j}{2}\cos\left[\frac{\pi}{2}\left(\frac{\bar{x}_j - \min(\bar{X})}{\max(\bar{X}) - \min(\bar{X})}\right)\right] - \sin\frac{\theta_j}{2}\sin\left[\frac{\pi}{2}\left(\frac{\bar{x}_j - \min(\bar{X})}{\max(\bar{X}) - \min(\bar{X})}\right)\right] \\ \sin\frac{\theta_j}{2}\cos\left[\frac{\pi}{2}\left(\frac{\bar{x}_j - \min(\bar{X})}{\max(\bar{X}) - \min(\bar{X})}\right)\right] + \cos\frac{\theta_j}{2}\sin\left[\frac{\pi}{2}\left(\frac{\bar{x}_j - \min(\bar{X})}{\max(\bar{X}) - \min(\bar{X})}\right)\right] \end{pmatrix} \\
&= \begin{pmatrix} \cos\left[\left(\frac{\pi}{2}\left(\frac{\bar{x}_j - \min(\bar{X})}{\max(\bar{X}) - \min(\bar{X})}\right)\right) + \frac{\theta_j}{2}\right] \\ \sin\left[\frac{\pi}{2}\left(\frac{\bar{x}_j - \min(\bar{X})}{\max(\bar{X}) - \min(\bar{X})}\right) + \frac{\theta_j}{2}\right] \end{pmatrix} = \begin{pmatrix} \cos\beta_j \\ \sin\beta_j \end{pmatrix}
\end{aligned} \tag{4}$$

where the phase $\beta_j$ of the quantum state $\left|\phi\left(\beta_j\right)\right\rangle$ is

$$\beta_j = \frac{\pi}{2}\left(\frac{\bar{x}_j - \min\left(\bar{X}\right)}{\max\left(\bar{X}\right) - \min\left(\bar{X}\right)}\right) + \frac{\theta_j}{2} \tag{5}$$

The values of phase $\theta_j$ are in $\left[0, \pi/2\right]$.

### *B. Deduction of the Phase Sum Formula*

Next, the quantum state $\left|\phi\left(\beta_1\right)\right\rangle$ is used as a control bit to control the controlled NOT gate to act on the quantum state $\left|\phi\left(\beta_2\right)\right\rangle$ to obtain a new quantum state $\left|\psi\left(\gamma_2\right)\right\rangle$.

The description of a controlled NOT gate is that only when all control bits are taken $\left|1\right\rangle$ is the output probability amplitude of the controlled bits reversed. If the control bits are taken as $\left|0\right\rangle$, the controlled values remain unchanged. The formula for controlling the NOT gate CNOT is expressed as follows:

$$\text{CNOT} = \begin{pmatrix} 1\,0\,0\,0 \\ 0\,1\,0\,0 \\ 0\,0\,0\,1 \\ 0\,0\,1\,0 \end{pmatrix} \tag{6}$$

First, the first controlled NOT gate is derived: $|\phi(\beta_1)\rangle = \begin{pmatrix} \cos\beta_1 \\ \sin\beta_1 \end{pmatrix}$ is the control bit, and $|\phi(\beta_2)\rangle = \begin{pmatrix} \cos\beta_2 \\ \sin\beta_2 \end{pmatrix}$ is the controlled bit; since the CNOT gate is a double qubit gate, it is necessary to first calculate the double qubit state $|\phi(\beta_1)\phi(\beta_2)\rangle$ of $|\phi(\beta_1)\rangle$ and $|\phi(\beta_2)\rangle$ to participate in the calculation:

$$|\phi(\beta_1)\phi(\beta_2)\rangle = |\phi(\beta_1)\rangle \otimes |\phi(\beta_2)\rangle = \begin{pmatrix} \cos\beta_1\cos\beta_2 \\ \cos\beta_1\sin\beta_2 \\ \sin\beta_1\cos\beta_2 \\ \sin\beta_1\sin\beta_2 \end{pmatrix} \tag{7}$$

The CNOT gate acts on a two-bit quantum state $|\phi(\beta_1)\phi(\beta_2)\rangle$, and its calculation result is as follows:

$$\begin{aligned} |\psi(\gamma_2)\rangle &= \text{CNOT}|\phi(\beta_1)\phi(\beta_2)\rangle \\ &= \begin{pmatrix} 1\,0\,0\,0 \\ 0\,1\,0\,0 \\ 0\,0\,0\,1 \\ 0\,0\,1\,0 \end{pmatrix}\begin{pmatrix} \cos\beta_1\cos\beta_2 \\ \cos\beta_1\sin\beta_2 \\ \sin\beta_1\cos\beta_2 \\ \sin\beta_1\sin\beta_2 \end{pmatrix} = \begin{pmatrix} \cos\beta_1\cos\beta_2 \\ \cos\beta_1\sin\beta_2 \\ \sin\beta_1\sin\beta_2 \\ \sin\beta_1\cos\beta_2 \end{pmatrix} \\ &= \cos\beta_1\cos\beta_2|00\rangle + \cos\beta_1\sin\beta_2|01\rangle \\ &\quad + \sin\beta_1\sin\beta_2|10\rangle + \sin\beta_1\cos\beta_2|11\rangle \end{aligned} \tag{8}$$

In the calculation result of formula (8), the probability amplitudes of the quantum states $|00\rangle$ and $|10\rangle$ are $\cos\beta_1\cos\beta_2$ and $\sin\beta_1\sin\beta_2$, respectively, and the probability amplitudes of the quantum states $|01\rangle$ and $|11\rangle$ are $\cos\beta_1\sin\beta_2$ and $\sin\beta_1\cos\beta_2$, respectively. Therefore, it can be inferred that the probability of the control result $|\phi(\gamma_1)\rangle$ of the controlled bit after passing through the controlled NOT gate in the basis state $|0\rangle$ is $\cos^2\beta_1\cos^2\beta_2 + \sin^2\beta_1\sin^2\beta_2$, and the probability of the basis state $|1\rangle$ is $\cos^2\beta_1\sin^2\beta_2 + \sin^2\beta_1\cos^2\beta_2$, which can be expressed as follows:

$$|\psi(\gamma_2)\rangle = \begin{pmatrix} \sqrt{\cos^2\beta_1\cos^2\beta_2 + \sin^2\beta_1\sin^2\beta_2} \\ \sqrt{\cos^2\beta_1\sin^2\beta_2 + \sin^2\beta_1\cos^2\beta_2} \end{pmatrix} = \begin{pmatrix} \cos\gamma_2 \\ \sin\gamma_2 \end{pmatrix} \tag{9}$$

in where

$$\begin{aligned} \gamma_2 &= \arcsin\left(\sqrt{\cos^2\beta_1\sin^2\beta_2 + \sin^2\beta_1\cos^2\beta_2}\right) \\ &= \arcsin\left(\sqrt{\frac{1}{4}\left[\sin(\beta_1+\beta_2) - \sin(\beta_1-\beta_2)\right]^2 + \frac{1}{4}\left[\sin(\beta_1+\beta_2) + \sin(\beta_1-\beta_2)\right]^2}\right) \\ &= \arcsin\left(\sqrt{\frac{1}{2}\sin^2(\beta_1+\beta_2) + \frac{1}{2}\sin^2(\beta_1-\beta_2)}\right) \end{aligned} \tag{10}$$

After the CNOT gate is activated, the quantum state $|\phi(\beta_1)\rangle$ is represented by a quantum state $|\psi(\gamma_1)\rangle$, and the phase angles of the two quantum states are equal:

$$\gamma_1 = \beta_1 \tag{11}$$

Therefore, the formula for the phase angle $\gamma_2$ can also be written as follows:

$$\gamma_2 = \arcsin\left(\sqrt{\frac{1}{2}\sin^2(\gamma_1+\beta_2) + \frac{1}{2}\sin^2(\gamma_1-\beta_2)}\right) \tag{12}$$

In this way, the expression for the phase of the quantum state $|\psi(\gamma_j)\rangle$ ($j$=2,3..., 35) can be obtained as follows:

$$\gamma_j = \arcsin\left(\sqrt{\frac{1}{2}\sin^2(\gamma_{j-1}+\beta_j) + \frac{1}{2}\sin^2(\gamma_{j-1}-\beta_j)}\right) \tag{13}$$

After the quantum state $|\psi(\gamma_{35})\rangle$ is obtained, it is used as a control bit to control the CNOT gate to act on the quantum state $|\psi(\gamma_1)\rangle$, and the formula for the output quantum state $|Y(\alpha)\rangle$ is as follows:

$$|Y(\alpha)\rangle = \begin{pmatrix} \sqrt{\cos^2\gamma_{35}\cos^2\gamma_1 + \sin^2\gamma_{35}\sin^2\gamma_1} \\ \sqrt{\cos^2\gamma_{35}\sin^2\gamma_1 + \sin^2\gamma_{35}\cos^2\gamma_1} \end{pmatrix} = \begin{pmatrix} \cos\alpha \\ \sin\alpha \end{pmatrix} \tag{14}$$

where

$$\alpha = \arcsin\left(\sqrt{\frac{1}{2}\sin^2(\gamma_{35}+\gamma_1) + \frac{1}{2}\sin^2(\gamma_{35}-\gamma_1)}\right) \tag{15}$$

The quantum state $|Y(\alpha)\rangle$ preserves the sum of the phases of 35 qubits from $|\phi(\beta_1)\rangle$ to $|\phi(\beta_{35})\rangle$. By measuring the quantum state $|Y(\alpha)\rangle$, the phase $\alpha_1$ can be obtained, and then the calculated value $\sin^2\alpha_1$ can be used to obtain the probability of the quantum state $|Y(\alpha)\rangle$ in the quantum basis state $|1\rangle$ as the real output y of the quantum manipulator, which is the square of the probability amplitude in the basis state in formula (14):

$$\begin{aligned} y &= \sin^2\alpha \\ &= \sin^2\left(\arcsin\left(\sqrt{\frac{1}{2}\sin^2(\gamma_{35}+\gamma_1) + \frac{1}{2}\sin^2(\gamma_{35}-\gamma_1)}\right)\right) \\ &= \left(\sqrt{\frac{1}{2}\sin^2(\gamma_{35}+\gamma_1) + \frac{1}{2}\sin^2(\gamma_{35}-\gamma_1)}\right)^2 \\ &= \frac{1}{2}\sin^2(\gamma_{35}+\gamma_1) + \frac{1}{2}\sin^2(\gamma_{35}-\gamma_1) \end{aligned} \tag{16}$$

where the values of $\gamma_1$ and $\gamma_{35}$ are shown in formulas (11) and (13), respectively.

According to equation (16), the expression for the real output y of the network can be obtained by performing polynomial operations on phases $\gamma_1$ and $\gamma_{35}$. According to equations (11) and (13), the expression for phase $\gamma_j$ can be obtained by performing polynomial operations on phases $\gamma_{j-1}$ and $\beta_j$, and $\gamma_1=\beta_1$ holds. Finally, according to equation (5), the expression for phase $\beta_j$ can be obtained by performing polynomial operations on a network that includes real input $x_j$ and weighted phase $\theta_j$.

In summary, the expression for the real output *y* of the network can be derived by polynomial operations of real input $x_j$ and weighted phase $\theta_j$, *j*=1,2,…,35, instead of using projection measurements, which requires $2^{35}$ exponential operations to solve the 35-qubit operations, thus solving the "exponential wall" problem.

## V. DEDUCTION OF NETWORK WEIGHT PHASE TRAINING FORMULA

In this section, the learning and training formulas for the 35 adjustable weight phases $\theta_j$, *j*=1, 2, ..., 35, are derived for the multi-input/single-output QPNN proposed in this chapter on the basis of the input–output relationship equation (16). The error *e* between the expected and actual outputs of each quantum manipulator is defined as follows:

$$e=t-y \tag{17}$$

where *e* is the network error, *t* is the expected output, and *y* is the actual output of the *i*-th quantum manipulator.

According to the gradient descent method, the formula for defining the output error function *E* is as follows:

$$E=\frac{1}{2}e^2 \tag{18}$$

The adaptive learning rate is used in the learning and training algorithm, which means that the value of the learning rate varies with the size of the error. The formula is as follows:

$$\eta^{(l+1)}=\begin{cases}1.05\eta^{(l)},\ \mathrm{SSE}^{(l)}<\mathrm{SSE}^{(l-1)}\\ 0.7\eta^{(l)},\ \mathrm{SSE}^{(l)}>1.04\,\mathrm{SSE}^{(l-1)}\\ \eta^{(l)},\ \text{others}\end{cases} \tag{19}$$

where the superscript *l* represents the current iteration number of learning and training, $\mathrm{SSE}=\sum e^2$, and the initial learning rate $\eta^1=\pi/4$.

All the learning rate values $\eta$ in the following formulas represent the values calculated in $l+1$ time, according to equation (19) in the $\eta^{(l+1)}$ iteration calculation.

Because $\theta_1$ first appears in expression (5) for the phase $\beta_1$ of quantum states $\left|\phi(\beta_1)\right\rangle$, and $\left|\phi(\beta_1)\right\rangle$ is equivalent to $\left|\psi(\gamma_1)\right\rangle$, while $\left|\phi(\beta_1)\right\rangle$ and $\left|\psi(\gamma_1)\right\rangle$ are involved in the phase summation of the first and the 35th CNOT gates, respectively, $\Delta\theta_1$ needs to be calculated separately. According to the gradient descent method, its calculation formula is derived as follows:

$$\Delta\theta_1=-\eta\times\frac{\partial E}{\partial\theta_1}=-\eta\times\frac{\partial E}{\partial e}\frac{\partial e}{\partial y}\frac{\partial y}{\partial\theta_1} \tag{20}$$

Substituting formula (16) into formula (20) for calculation yields:

$$\begin{aligned}\Delta\theta_1=-\eta\times e\times(-1)\times\frac{1}{2}\times\sin\left\{4\left[\frac{\pi}{2}\left(\frac{\bar{x}_1-\min(\bar{X})}{\max(\bar{X})-\min(\bar{X})}\right)+\frac{\theta_1}{2}\right]\right\}\\ \times\prod_{j=2}^{35}\cos\left\{2\left[\frac{\pi}{2}\left(\frac{\bar{x}_j-\min(\bar{X})}{\max(\bar{X})-\min(\bar{X})}\right)+\frac{\theta_j}{2}\right]\right\}\end{aligned} \tag{21}$$

Similarly, when *j* = 2, 3, ... , 35, $\theta_j$ first appears in expression (5) for the phase $\beta_j$ of the quantum state $\left|\phi(\beta_j)\right\rangle$, and each $\left|\phi(\beta_j)\right\rangle$ only participates in the phase summation of the *j*-th CNOT gate. Therefore, according to equation (21), the calculation formula of $\Delta\theta_j$ is as follows:

$$\Delta\theta_j=-\eta\times\frac{\partial E}{\partial e}\frac{\partial e}{\partial y}\frac{\partial y}{\partial\theta_j} \tag{22}$$

Substituting formula (16) into formula (22) for calculation yields:

$$\Delta\theta_j=-\eta\times e\times(-1)\times\frac{1}{2}\times\tan\left\{2\left[\frac{\pi}{2}\left(\frac{\bar{x}_j-\min(\bar{X})}{\max(\bar{X})-\min(\bar{X})}\right)+\frac{\theta_j}{2}\right]\right\}\times(1-2y) \tag{23}$$

In summary, the learning and training formulas for 35 adjustable weight phase values can be obtained as follows:

1) When *j*=1:

$$\begin{aligned}\Delta\theta_j=\eta\times e\times\frac{1}{2}\times\sin\left[2\pi\left(\frac{\bar{x}_1-\min(\bar{X})}{\max(\bar{X})-\min(\bar{X})}\right)+2\theta_1\right]\\ \times\prod_{j=2}^{35}\cos\left[\pi\left(\frac{\bar{x}_j-\min(\bar{X})}{\max(\bar{X})-\min(\bar{X})}\right)+\theta_j\right]\end{aligned} \tag{24}$$

2) when *j*=2, 3 , … , 35:

$$\Delta\theta_j=\eta\times e\times\frac{1}{2}\times\tan\left[\pi\left(\frac{\bar{x}_j-\min(\bar{X})}{\max(\bar{X})-\min(\bar{X})}\right)+\theta_j\right]\times(1-2y) \tag{25}$$

## VI. SIMULATION EXPERIMENT OF 2-LETTER RECOGNITION AND EXPERIMENTAL VERIFICATION ON THE QISKIT PLATFORM

### *A. Experimental Results*

1) Input the variable letters A and B as a $7\times5$ matrix composed of 0 or 1 as:

| input_matA= | input_matB= |
|---|---|
| [0 0 1 0 0 ; | [1 1 1 1 0 ; |
| 0 1 0 1 0 ; | 1 0 0 0 1 ; |
| 0 1 0 1 0 ; | 1 0 0 0 1 ; |
| 1 0 0 0 1 ; | 1 1 1 1 0 ; |
| 1 1 1 1 1 ; | 1 0 0 0 1 ; |
| 1 0 0 0 1 ; | 1 0 0 0 1 ; |
| 1 0 0 0 1 ]$^{T}$; | 1 1 1 1 0 ]$^{T}$; |

Straighten the pixelated matrix to obtain a $7 \times 5 = 35 \times 1$ vector, which is then used as input data for training. input_data = [input_matA(:), input_matB(:)]; there are:

input_data= [0 0 1 0 0 0 1 0 1 0 0 1 0 1 0 1 0 0 0 1 1 1 1 1 1 1 0 0 0 1 1 0 0 0 1 ;
1 1 1 1 0 1 0 0 0 1 1 0 0 0 1 1 1 1 1 0 1 0 0 0 1 1 0 0 0 1 1 1 1 1 0 $]^T$;

2) After normalization and polynomial phase encoding of the input data consisting of 0 and 1 for a given "A" and "B",

After training, a set of training solutions with a total error function of 0.0008 can be found. By substituting the trained weighted phase into the quantum circuit built by Qiskit, an approximate solution can be obtained. The maximum error between the Qiskit verification results and the MATLAB training results is only 0.0018. The analysis reveals that the output obtained from training is not precise to 0 and 1; it is only an approximate solution. Therefore, Qiskit's measurement results may also have deviations. This error can be reduced by increasing the Qiskit measurement frequency.

Through MATLAB and Qiskit experiments, it has been verified that a set of exact solutions can be found with an error function of 0 that satisfies the expected output weight phase. $[\theta_1, \theta_2, \theta_3, \ldots, \theta_{35}]$ The conditions that need to be met are as follows:

$$\theta_j = k_j \pi \tag{26}$$

where, $j$=1, 2, 3, ... , 35, $k_j \in \mathbb{Z}$.

### *B. Implementation and Results of the Qiskit Platform*

On the Qiskit platform, corresponding quantum circuits were constructed on the basis of the QPNN in Figure 2, as shown in Figure 3. To clearly display the circuit structure and highlight key annotations, the middle 28 quantum bits were omitted in Figure 3.

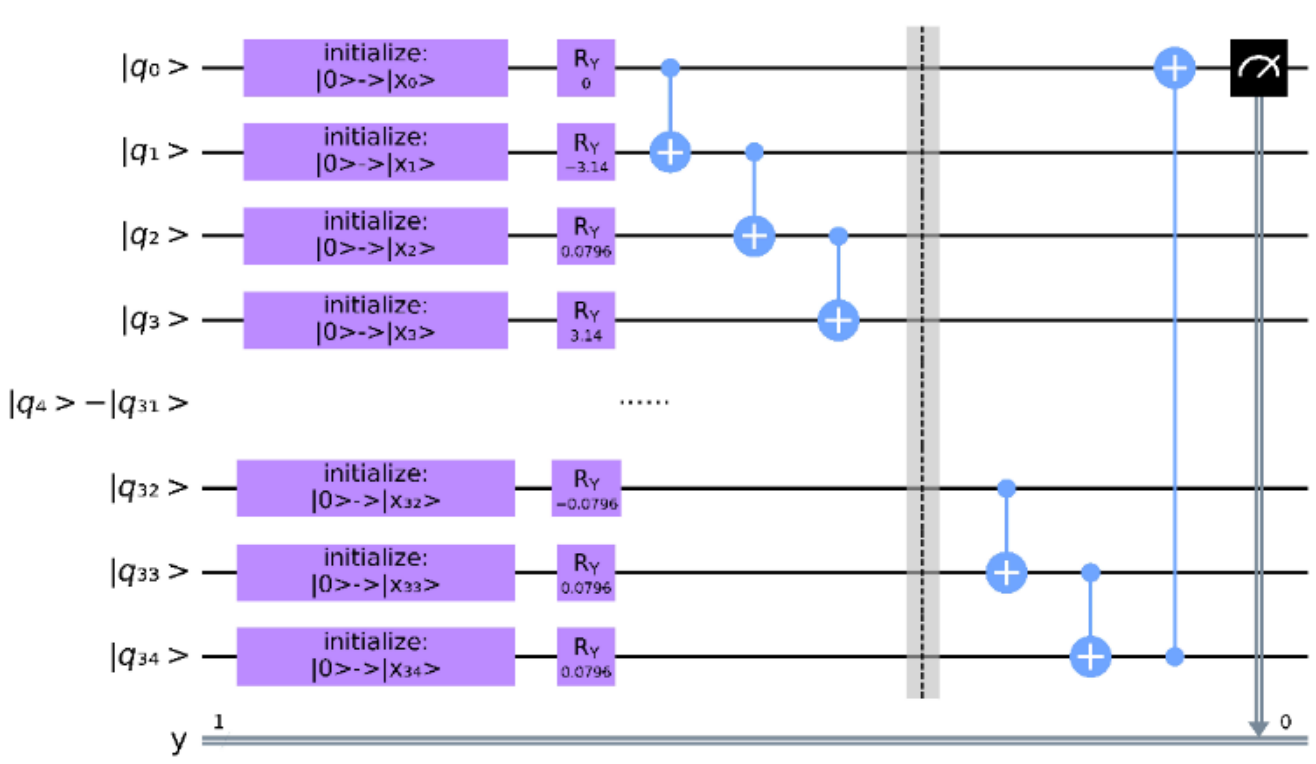

Fig. 4 Quantum circuit implemented on the Qiskit platform

The verification experiment results show that the successful recognition of letters' A 'and' B 'obtained on the basis of the designed QPNN structure diagram and the derived weight phase training formula is completely correct. The total error function of the quantum circuit verification result is 0.0016, which is essentially consistent with the accuracy of the total error function trained in MATLAB.

The actual measurement result of the quantum circuit constructed by Qiskit has a slightly higher total error function error than the MATLAB training result because of the statistical fluctuation caused by finite measurements; that is, the small random variation in the system near the equilibrium state. This random variation can be reduced by increasing the number of measurements so that the actual maximum output error is close to the expected maximum error.

On the basis of the designed QPNN structure diagram and the QPNN quantum circuit diagram implemented on the Qiskit platform, combined with design experience to set different initial phases, a set of weighted phase solutions for successfully identifying letters 'A' and 'B' with zero error in both the MATLAB environment and the Qiskit platform were obtained, as shown in equation (26). Due to the phase to $\pi$ value error on the actual experimental platform, there may be $10^{-4}$ error in the verification results.

## VII. CONCLUSIONS

Based on the proposed QPP/QPN [6] and [7], this paper designed a QPNN with 35 inputs/single output. The probability value of the network output for a single bit is calculated by the square of the sin angle of that bit, which is equal to the polynomial calculation results of various angles obtained after the action of different phase rotation gates and CNOT operations. The QPNN designed in this paper has a completely different structure from the existing standard QPNNs, which rely on repeated projection sampling for output. In order to distinguish QPNN proposed in this paper from traditional QPNNs, the QNN N. proposed in this paper is referred to as Cong-QPNN. Researches on Cong-QPNN with multi-inputs/ multi-outputs are widely underway, and more Cong-QPNN theories and their applications will soon emerge.